\documentstyle[preprint,pre,aps]{revtex}
\tighten
\begin{document}
\draft

\title{Estimation of the order parameter
exponent of critical cellular automata using
the enhanced coherent anomaly method}

\author{G\'eza \'Odor}

\address
{\it Research Institute for Materials Science, P.O.Box 49,
H-1525 Budapest, Hungary }

\date{\today}
\maketitle
\begin{abstract}

The stochastic cellular automaton of Rule 18 defined by Wolfram
[Rev.\ Mod.\ Phys.\ {\bf 55}, 601 (1983)] has been investigated
by the enhanced coherent anomaly method. Reliable estimate was
found for the $\beta$ critical exponent, based on moderate sized
($n \le 7$) clusters.
\end{abstract}

\pacs{05.40.+j, 64.60.-i}
\narrowtext

Calculating critical exponents of second order phase
transitions is a challenging problem. For nonequlibrium
systems, generalization of equilibrium statistical
physics methods is under developement. Among the most
notable analytical tools are series expansion \cite{dickj},
transfer matrix diagonalization \cite{kinzel} and the mean-field
renormalization-group method \cite{debell}.

In a series of earlier papers \cite{szab,odo,odoboc},
we have shown how the generalization of the mean-field technique
with appropriate extrapolation can be used to describe the critical
properties of cellular automata (CA) phase transitions.

The generalized mean-field approximation (GMF) first proposed
for dynamical systems by Gutowitcz et al.\cite{gut} and Dickman
\cite{dick} is shown to converge slowly at criticality.
In this method we set up equations for the steady state of the
system based on $n$-point block probabilities. Correlations with
a range greater than $n$ are neglected.
By increasing $n$ from 1 (traditional mean-field) step by step we
take into account more and more correlations and get better
approximations. The GMF approximation can be used as
the basis of the coherent anomaly method (CAM) calculation,
and it gives a reasonably good $\beta$ exponent for a dynamical
system with large ($n > 10$) \cite{inui}.
In this letter I show how an improved version of the
CAM proposed very recently \cite{kol} works on cellular
automata.

The essence of the CAM \cite{suz} is that the solution for
singular quantities at a given ($n$) level of approximation ($Q_n(p)$)
in the vicinity of the critical point is the product of the
classical singular behavior multiplied by an anomaly factor ($a(n)$),
which becomes anomalously large as $n \to \infty$ (and $p_c^n \to p_c$):
\begin{equation}
Q_n \sim a(n) (p/p_c^n - 1)^{\omega_{cl}}
\end{equation}
where $p$ is the control parameter and $\omega_{cl}$
is the classical critical index.
The divergence of this anomaly factor scales as
\begin{equation}
a(n) \sim (p_c^n - p_c)^{\omega - \omega_{cl}}
\end{equation}
thereby permitting the estimation of the true critical
exponent $\omega$, given a set of GMF approximation solutions.
However such an estimation depends to some extent on the
choice of the independent parameter ($p \leftrightarrow 1/p$).
To avoid this a corrected CAM was proposed \cite{kol}, based on
a new parameter:
\begin{equation}
\delta_n = (p_c/p_c^n)^{1/2} - (p_c^n/p_c)^{1/2} \label{del}
\end{equation}
such that Eq. (\ref{del}) is invariant under $p\leftrightarrow
p^{-1}$. This parametrization gives better estimates for the
critical exponents of the 3-dimensional Ising model \cite{kol}.

My target system for this kind of calculation was the
one-dimensional, stochastic Rule 18 CA \cite{wolf}.
This range-one cellular automaton rule generates a $'1'$
at time $t$ only when the right or the left neighbor
was $'1'$ at $t-1$  :
\begin{verbatim}

                    t-1 :           100          001
                    t :              1            1

\end{verbatim}
with probability $p$. In any other case the cell becomes $'0'$ at
time $t$. The order parameter is the concentration ($c$) of $'1'$-s.
For $p < p_c$ the system evolves to an absorbing state ($c=0$).
For $p\ge p_c$ a finite concentration steady state appears with a
continuous phase transition. This transition is known to belong
to the universality class of directed percolation (DP) \cite{cardy}.
At $t \to \infty$ the steady state can be built up from $'00'$ and $'01'$
blocks \cite{elo}. This permits one to map it onto stochastic
Rule 6/16 CA with the new variables $'01' \to '1'$ and $'00' \to '0'$:
\begin{verbatim}

          t-1 :           0 0          0 1          1 0        1 1
          t :              0            1            1          0

\end{verbatim}
and the GMF equations can be set up by means of pair variables.
In an earlier work \cite{szab} this was performed up to the order $n=6$,
and Pad\'e extrapolation was applied to the results.
Our best estimate for critical data was $p_c = 0.7986$
and $\beta = 0.29$.

Now, I have extended the GMF calculations up to $n=7$
(see Table \ref{table1}) with the help of the symbolic
Mathematica software. This required the setting up and solution
of a set of nonlinear equations of 72 variables. I obtained
$p_c^7 = 0.7729$, which is still $5 \%$ off the result obtained
by steady state simulation: $p_c = 0.8086(2)$ \cite{boc} or
from the more accurate time dependent simulation data:
$p_c = 0.8094(2)$ \cite{unp}.

The CAM analysis of $(a(n),\delta_n)$ data was done, taking
into account the correction term
\begin{equation}
a(n) = b \ \delta_n^{\beta - \beta_{cl}} +
       c \ \delta_n^{\beta - \beta_{cl} + 1}
\label{corr}
\end{equation}
and examining the stability of the solution by omitting different points
from the ($n = 1,..,7$) data set. For the fitting $\beta_{cl} = 1$
and $p_c = 0.8094$ were used. As was pointed out in Ref. \cite{kol} the
CAM data may contain departures from ideal scaling;
moreover there is no clear dependence on the order of the approximations.
I found relatively stable estimates using the correction formula
(\ref{corr}) on the data set with the omission of the $n=3$ point. The
$n=3$ approximation result does not fit into the $(\log(\delta_n) -
\log(a(n))$ curve either (see Fig. \ref{fig1}). Table \ref{table2}
shows the stability of the results, with the mean $\beta = 0.2796(2)$
calculated from them.
This compares very well with the value  $\beta = 0.2769(2)$
obtained by Dickman and Jensen \cite{dickj}
from series expansion. If the CAM calculation
based on $p$ or $1/p$ independent variables the results
differ by $\pm 0.005$ from the present enhanced version $\beta$ estimates.

Another critical model with non-DP universality, the
nonequilibrium kinetic Ising model, has been examined with the
enhanced CAM method, and the $\beta$ exponent estimate is in
agreement with the simulation results \cite{mo}.

The conclusion of this study is that the enhanced version CAM
method with careful data analysis gives good estimates for
the critical exponent for moderate $n < 10$ level GMF approximations.
Calculation of the $n=5,6,7...$ \ GMF approximations is possible
on moderate sized workstations. The solution of the $n=7$ level
approximation took about 10 hours CPU time on a SUN Sparc-10.
This provides an efficient analytical tool for
exploring universalities of nonequlibrium systems.

This research was partially supported by the
Hungarian National Research Fund (OTKA) under grant numbers
T-4012 and F-7240.

\figure{Result obtaining by applying the improved CAM method
on $n$-pair ($n = 1,..,7$) approximation data.
The logarithm of the anomaly coefficient $a(n)$
is plotted versus the logarithm of the improved independent
variable $\delta_n$.
Fitting was done according to Eq.\ (\ref{corr}).
\label{fig1}
}

\narrowtext
\begin{table}
\caption{GMF calculation results for pair approximation data}
\begin{tabular}{lrl}
$n$ & $p_c^n$   & $a(n)$ \\
\tableline
$1$ & $0.5000$  & $0.5000$  \\
$2$ & $0.6666$  & $1.5000$  \\
$3$ & $0.7094$  & $2.3484$  \\
$4$ & $0.7413$  & $2.8816$  \\
$5$ & $0.7543$  & $3.5345$  \\
$6$ & $0.7656$  & $4.2545$  \\
$7$ & $0.7729$  & $4.8463$  \\
\end{tabular}
\label{table1}
\end{table}

\narrowtext
\begin{table}
\caption{CAM calculation results for pair approximation data}
\begin{tabular}{lrl}
data  & $\beta$ \\
\tableline
$1-2  -4-5    $& $0.273$ \\
$1-2  -4-5-6  $& $0.271$ \\
$1-2  -4-5-6-7$& $0.282$ \\
$1-2  -4-5  -7$& $0.285$ \\
$1-2  -4  -6-7$& $0.275$ \\
$1-2    -5-6-7$& $0.310$ \\
$1    -4-5-6-7$& $0.275$ \\
$2  -4-5-6-7  $& $0.266$ \\
mean           & $0.2796(2)$ \\
\tableline
Pad\'e extrapolation, ref.\cite{szab}&$0.29$\\
simulation, ref.\cite{boc}&$0.285(5)$\\
series expansion for DP ref.\cite{dickj}&$0.2769(2)$
\end{tabular}
\label{table2}
\end{table}

\end{document}